\documentstyle{mn}
\begin{document}
\bibliographystyle{plain}
\title[Evolution of cooling flow]
           {Cooling Flows Induced by Compton Cooling due to Luminous Quasars
in Clusters of Galaxies}
\author[NISIKAWA A. HABE A. ISIBASI N.]{NISIKAWA Atuko,$^1$
\thanks{E-mail:akko@phys.hokudai.ac.jp}
Asao HABE,$^1$
\thanks{E-mail:habe@phys.hokudai.ac.jp}
and
ISIBASI Nobuo$^2$
\thanks{E-mail:nobuo@nerv.ed6.info.kanagawa-u.ac.jp}\\
$^1$Department of Physics, Hokkaido University, Kita-ku,
Sapporo 060 , Japan\\
$^2$Department of Information Science, Kanagawa University, Tutiya,
Hiratuka 259-12, Japan\\}
\date{ ( Submitted   \qquad      )}
\maketitle
\begin{abstract}
We have studied the effects of Compton cooling on cooling flow by
performing numerical hydrodynamic calculations of the time evolution
of hot gas in clusters of galaxies with luminous quasars.
We assumed various temperatures for the hot intracluster gas.
We have shown that the Compton cooling due to very luminous quasar is
effective in inducing cooling flow,
before radiative cooling flow is realized.
The mass flux due to the cooling flow increases with time,
as the Compton cooled region expands.
However, the mass flux of the Compton cooling flow is not large, less than
1 $M_\odot $ yr $^{-1}$ in our model,
since Compton cooled region is limited in an inner galactic region around
a quasar.
Even though the quasar active phase ceased, the cooling flow will continue
for at least $10^9$ yr.
The accreted mass is enough to explain X-ray absorption lines in high red
shift quasars,
if the Compton cooled gas is compressed by high pressure intracluster gas.

\end{abstract}
\section{INTRODUCTION}
Radio-loud quasars have some  characteristic features different  
from radio-quiet quasars.
Radio-loud quasars are observed to be strong UV sources 
($L_{\rm{UV}} \sim 10^{46} - 10^{47}$ erg s $^{-1}$) 
(e.g., McDowell {\it et al.} 1989),
and  host galaxies of  radio-loud quasars are found preferentially
in high density environments (e.g., Smith and  Heckman 1990)
and are often  
central elliptical galaxies in clusters (Malkan 1984).
Cooling flows sometimes appear around these quasars
(Fabian 1994).
For examples, Elvis {\it et al.} \shortcite{El} have shown strong X-ray 
absorption lines in high redshift quasars.
They have suggested that the origin of absorbing gas 
is a cooling flow around quasar and
have estimated that the column density of gas that absorbs
the X-ray is up to $10^{23}$ cm$^{-3}$.
Usually cooling flows in clusters of galaxies 
are proposed to be accretion flows
of intracluster medium induced by radiative cooling
(Sarazin 1986; Fabian {\it et al.} 1987; Forman 1988).
Since radiative cooling time may be longer than the cosmic age at such a
high $z$,
another cooling process is needed for the cooling flow around quasar.
Authors suggested that
cooling flows are induced by Compton cooling due to strong
UV photons of quasars.
\par
For the majority of quasars the bulk of the emitted electromagnetic
luminosity is observed in the optical/UV/soft-X-ray spectral domain,
the so-called big blue bump (Shields 1978; Malkan 1983).
The origin of the big blue bump is accretion through a Keplerian disk or a
non-Keplerian torus onto a supermassive black hole (Rees 1984).
Fabian and Crawford \shortcite{FC}
(hereafter F.C.) have proposed that,
since Compton scattering between strong UV photons and hot electrons
can dominate cooling process for the hot gas
in central region of a cluster,
accretion by Compton cooling
can supply the gas to a central massive black hole,
and have shown that
the accretion rate by Compton cooling flow can be larger than
that of Bondi-accretion flow.
We summarise this point according to F.C. as follows.
\par
The Bondi accretion rate, $\dot{M_{\rm B}}$, may be estimated as
\begin{equation}
\label{Mb}
 \dot{M_{\rm B}} = \frac{4\pi \alpha ^2}{c_{\rm s}^3} (GM)^2 \rho,
\end{equation}
where $\rho $, $c_{\rm s}$, M, $\alpha $, are the gas density,
sound speed, the central black hole mass
and a factor depending on the adiabatic index of the gas,
respectively.
We estimate the typical value of $\dot{M_{\rm B}}$ in equation (\ref{Mb}) as,
$\dot{ M_{\rm b}} \sim 2.0 \times 10^{-3}$  $M_{\odot}$ yr$^{-1}$, 
where we take $\rho = 6 \times 10^{-26}$ g cm$^{-3}$, 
temperature $T = 10^7$ K, $M = 10^9$  $M_9$  $M_{\odot}$
and $\alpha = 1$.
At a radius $R$, the time-scale by Compton cooling, $\tau_{\rm c}$, is
\begin{equation}
  \tau _{\rm c} \approx 10^{13} \; R_2^{2} \; L_{47}^{-1} \; \rm s,
\end{equation}
where the radius $R=100 \; R_2 $ pc and
      the luminosity $L = 10^{47}$ $L_{47}$ erg s$^{-1}$.
While the time-scale by bremsstrahlung cooling, $\tau_{{\rm rad}}$, is
\begin{equation}
  \tau_{{\rm rad}} \approx 6 \times 10^{13} \; T_7^{3/2} \; P_8^{-1} \; \rm s,
\end{equation}
where the temperature $T = 10^7$ $T_7$ K
and the pressure $P = nT = 10^8$  $P_8$ cm$^{-3}$ K.
Therefore within a radius
\begin{equation}
  R_{\rm c} \approx 250 \; L_{47}^{1/2} \; T_7^{3/4} \; P_8^{-1/2} \; \rm {pc},
\end{equation}
Compton cooling dominates bremsstrahlung cooling in the highly ionized
conditions, so that we can expect stable gas inflow
since the cooling time does not depend on gas density.
Consequently, all the gas at $R_{\rm c}$ flows into the central black hole, 
the accretion rate in this case is
\begin{eqnarray}
\label{Mc}
  \dot{M_{\rm c}}
  & \approx & \frac{4\pi R_{\rm c}^3 \rho} {\tau_{{\rm rad}}} \nonumber\\
  & \sim    & 1.45 \; L_{47}^{3/2} \; T_7^{1/4}
             ( \frac{\rho}{6 \times 10^{-26} \; {\rm g \; cm}^{-3}} )^{1/2} \;
             M_{\odot} \; \rm{yr}^{-1},
\end{eqnarray}
that is much larger than the Bondi accretion rate.
The dependence of $\dot{M_{\rm c}}$ on $L$, i.e.
$\dot{M_{\rm c}} \propto L^{3/2}$ and the linear
dependence of $L$ on $\dot M$, means that,  if $L > L_{\rm c}$,
the accretion induced by 
Compton cooling increases $L$  further,
where $L_{\rm c}$ is the quasar luminosity at 
the accretion rate of equation (\ref{Mc}).
On the other hand, if $L < L_{\rm c}$ then $\dot{M_{\rm c}}$
is reduced and the luminosity continues to decline to $L_{\rm B}$
that is defined by the luminosity of a quasar at the mass accretion rate of
$\dot{M_{\rm B}}$.
\par
When Compton cooling time is shorter than
radiative cooling time in the inner central region,
Compton cooling can induce the gas inflow
before the radiation cooling induces the gas inflow.
However, when the life time of quasar is comparable to
Compton cooling time of the gas within $R_{\rm c}$,
gas flow can not be steady state during the duration time
of the quasar.
Therefore we should study numerically the time evolution of the gas inflow
induced by Compton cooling.
Moreover we expect another possibility 
that by the density concentration due to Compton cooling
radiative cooling time would be shorter than Compton cooling time,
so that the hot gas would be cool radiatively at the central region
and that the density concentration could go on.
We examine whether the mass accretion rate increases
by the effect of Compton cooling
and how Compton cooling affects 
the evolution of the intracluster gas by using hydrodynamic scheme.
We fix the quasar luminosity in our model for simplicity
through our all simulations.
\par
In \S 2, we describe our model of the luminous quasar
with the cooling flow in the central galaxy in a cluster.
Our model incorporates the elliptical galaxy that includes stars, 
dark halos and cooling flow.
In \S 3, we describe the results of our model calculations.
In \S 4, we discuss the results and summarize our conclusions.

\section{A MODEL FOR COOLING FLOW WITH LUMINOUS QUASAR IN A CENTRAL CLUSTER
   GALAXY}
The central concept of our model is
the evolution of density, temperature and accretion rate
of the cooling flow in luminous elliptical host galaxy
with a quasar.
We assume that the age of quasar is $10^8$ yr.
By treating the quasar as a source of UV photons,
we consider Compton cooling process.
\par
Since we concentrate on the evolution of the intracluster gas near
the central galaxy, we calculate gas dynamics near the central galaxy.
We assume that the cooling flow is spherically symmetric for simplicity.
%
%
\subsection{Mass distribution of the stellar system and dark halo}
The stars in the central galaxy are assumed to be distributed
as the modified King profile with the density distribution given by
\begin{equation}
  \rho_*(r) = \rho_{\rm c} \; [ \: 1 + (r/r_{\rm c})^2 \: ]^{-3/2},
\end{equation}
where r is the radial distance and $r_{\rm c} = 500$ pc is the core radius
of the stellar distribution and $\rho_{\rm c} = 3.0 \times 10^{-21}$
g cm$^{-3}$.
Observations indicate that elliptical galaxies do not have flat cores, but 
rather cusp profiles (Hernquist 1990).
However, the mass inside cusp is not so large, if we compare the  mass
within the core region.
As we will show in Section 3,
the cooling flow flux is mainly determined by the difference of pressures  
between at the inside and at the outside of the cooled region. 
Therefore, the choice of mass model with or without cusp is not so serious
for our model.
\par
We have assumed that the dark halo
has a total mass which increases linearly with radial distance
as suggested for early-type galaxies by X-ray mass determinations
(e.g., Forman et al. 1985).
The particular dark matter density distribution
we have adopted is
\begin{equation}
  \rho_{\rm d}(r) = \rho_{\rm {dc}} \; [ \: 1 + (r/r_{\rm d})^2 \: ]^{-1},
\end{equation}
where $r_{\rm d} = 10$ kpc is the core radius of the dark halo distribution and
$\rho_{\rm {dc}} = 6.26 \times 10^{-25}$ g cm$^{-3}$.
\par
 The stellar and dark matter mass distributions, $M_*(r)$ and
$M_{\rm D}(r)$, can be determined by integrating the respective expressions
for the densities. Carrying out the integrations, the mass distributions
are found to be
\begin{equation}
 M_*(r) = 4\pi \: \rho_{\rm c} \: r_{\rm c}^3 \;
         [ \: \ln ( x + \sqrt{x^2 + 1})
                   -x/{\sqrt{x^2 + 1}} \: ]
\end{equation}
and 
\begin{equation}
 M_{\rm D}(r) = 4\pi \: \rho_{\rm {dc}} \: r_{\rm d}^3 \; (y - \tan^{-1}y),
\end{equation}
where $x=r/r_{\rm c}$ and $y=r/r_{\rm d}$.
The corresponding virial temperature of the model galaxy at $r_{\rm c}$
is $2 \times 10^7$ K.
\par
We consider a supermassive black hole (hereafter SMBH) at the centre of galaxy.
We assume that the mass of SMBH is $M_{\rm BH} = 10^9$ $M_{\odot}$.
The SMBH mass dominates  within the radius 
$$ r_{\rm in} =
   172 \; (\frac{{M_{\rm BH}}}{10^9 \; M_{\odot}})^{1/3} \; \rm {pc}. $$
At such a radius the virial temperature $T_{\rm BH }$ due to the SMBH is
$$ T_{\rm BH} =
   1.9 \times 10^6 \;
   (\frac{M_{\rm {BH}}}{10^9 \; M_{\odot}})^{2/3} \;
   \rm K.$$
%
%
\subsection{The initial conditions}
 Given the mass distribution of the stars and dark matter,
the initial density distribution of the hot gas can be derived by
assuming that it is in hydrostatic equilibrium. This assumption is
valid when flow velocity is small. This condition can be written
\begin{equation}
  \frac{dP(r)}{dr}=-\rho(r)\frac{GM(r)}{r^2},
\end{equation}
where $P(r)$ is the gas pressure, $\rho(r)$ is the gas density distribution,
and G is the gravitational constant. The total mass $M(r)$ is given by
\begin{equation}
  M(r) = M_*(r) + M_{\rm D}(r) +  M_{\rm BH}.
\end{equation}
The gas density distribution for an isothermal temperature
is given by
\begin{equation}
  \rho(r) = \rho_{\rm g0} \:
            \exp[ \: -(\phi(r) - \phi(0))\frac{\mu \, m_{\rm H}}{kT_0} \: ],
\end{equation}
where 
$k$ is the Boltzmann constant, $T_0$ is the initial isothermal temperature, 
and $\phi(r)$ is the gravitational potential by the stellar and
dark matter mass distributions.
We take $\rho_{\rm g0} = 6 \times 10^{-26} \; \rm {g \; cm^{-3}}$,
which maybe a typical value of interstellar matter in the host galaxy.
We assume various value of $T_0$ as shown in Table 1,
since we want to study Compton cooling effects on intracluster gas
in various types of cluster of galaxies.
We redefine $R_{\rm c}$ by using equation (2) and $\tau_{\rm c} = 10^8$ yr.
By the new definition, we get
$R_{\rm c} = 540$ pc for $L_{\rm {UV}} = 10^{46} \; {\rm erg \; s}^{-1}$ and
$R_{\rm c} = 1700$ pc for $L_{\rm {UV}} = 10^{47} \; {\rm erg \; s}^{-1}$.
\par
Since we assume that flow velocity is small
except for the vicinity of a SMBH, 
the initial velocity distribution in the central region is given by
\begin{equation}
 \label {vinner}
 v(r) =-\frac{n_{\rm e}(r) \, n_{\rm i}(r) \, \Lambda_0 \: T_0^\alpha \, r^2}
             {\rho(r) \, G \, M(r)},
\end{equation}
where $n_{\rm e}(r)$ is the electron density, $n_{\rm i}(r)$ is the
ion density, $\Lambda_0 \; T_0^\alpha$ is the emissivity.
We use the steady state energy equation.
In the outer region, the velocity distribution is given by
\begin{equation}
 v(r) =-\frac{r_{\rm inner}^2 \: \rho (r_{\rm inner}) \, v(r_{\rm inner})}
             {r^2 \, \rho (r)}.
\end{equation}
We assume that  $r_{\rm inner}$ is  $r_{\rm d}$.

We assume that cooling function $\Lambda (T) = \Lambda_0 \; T^\alpha$.
We take $\Lambda (T)$ to be that given by Raymond, Cox and Smith \shortcite{RCS}
for an optically thin, fully ionized gas with the cosmic abundance for $T
\gid 10^4$ K
and zero for $T < 10^4$ K.
$\Lambda (T)$
is given cgs unit by
\begin{equation}
   \Lambda(T) = \left \{
    \begin{array}{ll}
        2.2 \times 10^{-27} \: T^{0.5},   & \: \mbox{for $T > 4 \times
10^7$} \\
        5.3 \times 10^{-17} \: T^{-0.87}, & \: \mbox{for $4 \times 10^7 > T
> 2 \times 10^6$} \\
        6.2  \times 10^{-19} \: T^{-0.6}, & \: \mbox{for $2 \times 10^ 6
> T > 10^5$} \\
        1.0  \times 10^{-24} \: T^{0.55}, & \: \mbox{for $10^5 > T > 10^4$.}
    \end{array} \right. 
\end{equation}
%
%
\subsection{The basic equations}
We calculate the time evolution of
spherically symmetric X-ray emitting hot gas models.
We assume that the gas is ideal. 
The hydrodynamic equations for a hot gas are 
\begin{equation}
  \frac{\partial \rho}{\partial t}
 +\frac{1}{r^2}\frac{\partial }{\partial r}(r^2\rho v) = 0,
\end{equation}
\begin{equation}
  \frac{\partial \rho v}{\partial t}
 +\frac{1}{r^2}\frac{\partial }{\partial r}(r^2\rho v^2)
 =-\frac{\partial P}{\partial r} - \rho \frac{GM}{r^2},
\end{equation}
\begin{equation}
  \frac{\partial U}{\partial t}
 +\frac{1}{r^2}\frac{\partial }{\partial r}\{r^2(U + P) v\}
 =-\rho v\frac{GM}{r^2} - n_{\rm e}n_{\rm i}\Lambda_0T^\alpha
  -\frac{\rho \Lambda_c(T)}{r^2},
\end{equation}
where
\begin{equation}
  P=\frac{\rho \, k \, T}{\mu \, m_{\rm H}}
\end{equation}
is the gas pressure and
\begin{equation}
  U = \frac{1}{2}\rho v^2 + \frac{P}{\gamma - 1}
\end{equation}
is the sum of kinetic energy and thermal energy per unit volume, and
\begin{equation}
 \Lambda_{\rm c}(T) = \Lambda_{\rm {c0}} \: L \: T
\end{equation}
is the coefficient related with  the Compton cooling
and  $L $ is the UV luminosity of the quasar.
$\Lambda_{\rm {c0}}$ is the constant,
\begin{equation}
  \Lambda_{\rm {c0}} = \frac{1.1 \: \sigma_T \: k}
                            {2.1 \: \mu \: m_{\rm e} \: c^2},
\end{equation}
where $\sigma_{\rm T}$ is the cross section for the Thomson scattering.
We take the $\gamma = 5/3$.
Other physical variables have usual meanings.
%
%
\subsection{Timescale}
We have performed numerical calculations in the variety of cases.
In Table \ref{tab-init}, we show the numerical model parameters
and Bondi accretion rate $\dot{M_{\rm b}}$
to compare with our numerical results.
We take $\alpha = 1$ and $\tau_{\rm c} =10^8$ yr.
In Fig. \ref{fig-cool1}, we show the cooling time ($\tau_{\rm c}$
and $\tau_{\rm {rad}}$)
in the variety of temperature and luminosity.
It is interesting that Compton cooling time depends  on the radius
and luminosity and does not depend on the gas density and temperature.
%
%
\subsection{The simulation}
 We assume six cases of temperature of intracluster medium.
Instead of the gravitational confinement of the hot intracluster gas
in the cluster of galaxies,
we assume the outer boundary confines this hot gas.
 We use the flux-split scheme with second-order accuracy in space and
the first-order accuracy in time (van Leer 1982, van Albada,
van Leer and Roberts 1982 ).
This scheme is the one of the upwind scheme for the Euler equation.
We calculate the region from $r=50$ pc to $50$ kpc.
The calculated region covers the black hole mass dominated region
as shown in Section 2.1.
We use a mesh with $4996$ elements.
\par
We take the boundary conditions as follows.
At the inner and outer boundaries, the spatial gradient of the velocity
is zero.
At the inner boundaries, we assume that gas mass flux and   total energy
flux are
same those in the cell next to the inner boundary.
By this inner boundary condition, we obtain   smooth accretion flow.
At the outer boundaries,the density and energy are fixed in time.
\par
We have calculated the evolution of the cooling flow for  the both cases of
the existence of Compton cooling
and no Compton cooling (only radiative cooling)
for $t=1 \times 10^9$ yr from the initial situation.
\section{NUMERICAL RESULTS}
We have shown that Compton cooling is dominant cooling process within
$R_{\rm c}$
which is defined by equation (4)
and that there is the region where Compton cooling time is shorter than 
the age of a quasar as shown in Fig. 1.
In this region, we can expect a cooling flow induced by Compton cooling
within the lifetime of a quasar.
From our numerical results, we found that Compton cooling induces a cooling
flow,
even if the temperature of intracluster gas is as high as $10^8$ K.
However, the mass flux is smaller than that estimated by equation (5),
since the lifetime of a quasar is too short
to develop the cooling flow structure.
We found the case in which the  cooling flow induced by Compton cooling
continues after the lifetime of quasar in the lower temperature
intracluster gas case.
\par
In Fig. \ref{fig-flow1},
we show the results of 7.d747 model as the strong Compton
cooling flow case.
These figures show the evolution of the density,
the velocity, the temperature and the accretion rate of the gas in 7.d747
model .
In this model, we assume the luminous quasar with
$L_{\rm {UV}} = 10^{47}$ erg s$^{-1}$ and
the temperature of intracluster gas, $7 \times 10^7$ K.
In this case, since Compton cooling is very effective,
the gas in the inner region of galaxy cools by Compton cooling
within the lifetime of a quasar
and the cooled gas inflow is realized as shown in Fig. 2(b).
We should note that
radiative cooling time of the hot gas initially filled in the host galaxy
is longer than the calculation time and
we do not find any cooling flow in 7.d7rad model
in which we do not consider Compton cooling.
The inflow velocity attains $40$ km s$^{-1}$ at $r_{\rm c}$ at $10^8$ yr.
The mass flux is as large as 0.45 $M_{\odot} \; {\rm yr}^{-1}$ in this stage.
This is smaller than that estimated by equation (5).
After the lifetime of a quasar, we stop Compton cooling,
that is, we calculate the energy equation without Compton cooling.
After then, the cooled gas flows into the central region, and
the hot gas occupies the whole region.
Finally, the cooling flow disappears.
\par
Next, we show the weak Compton-cooling-flow case in Fig. 3.
In Fig. 3, we show 7.d746 model.
Fig. 3 shows that the temperature of gas
within 500 pc decreases due to Compton cooling.
As a result, the gas density increases by the compression
due to the high pressure of the outer hot gas.
However, the flow velocity and the mass flux
are very small in the inner region,
as shown in Fig. 3(b) and Fig. 3(d).
The hot gas in the inner region is in nearly hydrostatic equilibrium.
The small mass flux would not affect the evolution of a quasar and the host
galaxy.
Since the gas cooled by Compton cooling is not thermally unstable
as shown by equation (2)
and the gas temperature in this region  remains rather high temperature
in comparison with the virial temperature of the host galaxy,
the cooling flow is not realized.
This is the reason why gas is nearly hydrostatic state.
\par
Next, we show the Compton-cooling-induced-cooling flow case in Fig. 4.
Fig. 4 shows the numerical results of 5d747 model
and that the gas in the inner region rapidly cools by Compton cooling,
and attains less than $10^7$ K within $10^8$ yr in 
the central region of the host galaxy.
In this case, if we do not consider Compton cooling,
radiative cooling flow does not appear as shown in Fig. 5.
Since the temperature of gas near $r_{\rm c}$ becomes much less than
the virial temperature of the host galaxy, $2 \times 10^7$ K,
the gas inflow is induced and the mass flux attains to
0.4 $M_{\odot} \; {\rm yr}^{-1}$.
After $10^8 \; {\rm {yr}}^{-1}$, we stop Compton cooling,
and the inflow velocity decreases.
However, radiative cooling flow continues to $10^9$ yr.
The mass flux is 0.2 $M_{\odot} \; {\rm {yr}}^{-1}$ at $10^9$ yr.
We can conclude that the cooling flow after the lifetime of a quasar in 5d747
model is initially induced by Compton cooling and
is maintained by radiative cooling.
\par
Next, We show the numerical results of 5d746 model in Fig. 6.
In 5d746 model,
we assume one-tenth of a quasar luminosity of 5d747 model,
$10^46$ erg s$^{-1}$.
In this case,
the gas in the inner region of the host galaxy cools by Compton cooling and
the density in the cooled gas region increases.
However, the accretion rate is as small as 0.01 $M_{\odot} \; {\rm {yr}}^{-1}$,
since the Compton cooling rate is smaller than 5d747 model.
After Compton cooling stopped,
the gas density and temperature return to the initial state.
Although the gas cools by Compton cooling,
the cooling flow disappears after Compton cooling stops.
From these results,
we can interpret that the strong Compton cooling induces radiative cooling flow
in 5d747 model.
\section{DISCUSSION}
We have shown that, if a quasar has very strong UV source 
($L_{\rm {UV}} \gid 10^{47} \; \rm {erg \; s}^{-1}$),
the cooling flow can be induced by the  quasar during the quasar active phase.
In this case, intracluster gas is accumulated into the host galaxy.
The total mass accreted into the host galaxy is not so large 
but enough  mass to maintain the activity of  an active galactic nuclei.
On the other hand,
for the typical lifetime of a quasar, $10^8$ yr,
a quasar activity ceases before Compton cooling becomes effective 
to induce cooling flow for the high temperature
intracluster gas case (e.g., initial gas temperature of   $\sim 10^8$ K),
if a quasar does not have a strong UV source ($L_{\rm {UV}} \lid
10^{46}$ erg $\rm s^{-1}$).
\par
Due to Compton cooling, the gas temperature decreases,
and the gas density increases by the compression of high pressure of outer gas.
The gas flows in this stage corresponds to the pressure driven accretion
(Sarazin 1986).
If the temperature of cooled gas becomes lower than the virial temperature of 
the host galaxy, the inflow induced by Compton cooling occurs.
As shown in Table 2,  the mass flux at $10^8$ yr can be approximated 
by using the equation (5),
except for replacement of $\tau _{\rm {rad}}$ by $10^8$ yr and $R_{\rm c}$
by the distance where Compton cooling time equals to $10^8$ yr.
Then, equation (5) should be changed
as $\dot M_{\rm c} \sim 0.5 \; L_{47}^{3/2}$.
It is possible, that the feedback on central accretion
due to the increased quasar luminosity,
as Compton cooling become effective,
would increase the mass flow from large radii to large enough values
to account for central accretion rate,
if the quasar UV luminosity reaches $10^{48}$ erg s $^{-1}$.
\par
Compton cooling affects the halo gas in the host galaxy and 
increases its gas density.
As a result,
after a quasar activity ceases and Compton cooling become ineffective,
there are the cases that radiative cooling flow continues,
although the mass flux is smaller.
\par
Compton cooling is not thermally unstable,
since Compton cooling does not depend on the gas density.
Although Compton cooling time is much smaller than the simulation time,
we have gotten the results that the temperature of hot gas
does not attain very low temperature in the inner region
in the high temperature case.
The reason is as follows:
In 7d746 model, the flow time at $10^8$ yr is $\sim 10^7$ yr at 100 pc,
since $v \sim 10$ km $\rm s^{-1}$.
There is not enough time to cool further for Compton cooled gas
in the calculated region, since Compton cooling time in this region 
is as large as $10^7$ yr.
On the other hand, in 5d747 model,
the flow time at $10^8$ yr is as large as $4 \times 10^7$ yr
at $r \sim 1000$ pc, since $v \sim 25$ km s$^{-1}$.
Since Compton cooling time is as large as $10^7$ yr at $r \sim 1000$ pc, 
Compton cooling is effective to reduce the temperature of intracluster gas.
\par
Elvis {\it et al.} (1994) showed the evidence of X-ray absorption
in high redshift quasars.
They found that these quasars are very luminous
X-ray sources up to $10^{48}$ erg $\rm s^{-1}$.
They suggested that a cooling flow around the
quasar accounts for the X-ray absorption.
If these quasars have comparable magnitude of UV flux with X-ray,
Compton cooling would be very effective in these quasars.
As discussed above, since $\dot{M_{\rm c}} \propto {L_{\rm {UV}}}^{3/2}$,
the mass accretion rate becomes large.
This cooled gas in Compton cooling flow would correspond to X-ray absorption.
We compare column density of our Compton cooling flow models with
the column density
observed in the high redshift quasars.
As shown in Section 3, the gas density distribution may be approximated by
$\rho = \rho_0(r/r_0)^{-1}$,
where $\rho_0 \sim 2 \times 10^{-25} \; \rm{g \; cm}^{-3}$
and $r_0 \sim 1000$ pc.
The column density $N$ is estimated as
$$ N=10^{21}
     (\frac{\rho_0}{2 \times 10^{-25} \; {\rm g \; cm^{-3}}})
     (\frac{r_0}{1 \; \rm {kpc}})
  \ln \frac{(r_{\rm{max}}/{1 \; \rm{kpc}})}{(r_{\rm {in}}/{1 \; \rm {pc}})}
     \; \rm {cm}^{-3}. $$
This column density is smaller than the observed value.
The compression of cooled gas is expected,
since the pressure of central gas is very high.
It is possible to explain the high column density at quasar observed in
high redshift
by Compton cooling flow by this compression.
\par
In our numerical results, 
the cooled gas flows though the inner boundary.
It is not clear that how the gas accretes onto the central black hole or 
an accretion disk around the SMBH.
Then we hold our model luminosity fixed.
The cooling of inflow from large radii to accretion at small radii
is beyond the scope of our model.
A particular question is whether the increase in the virial temperature
inside 100 pc due to the presence of a SMBH leads to efficient infall
to smallse radii or not.
\par
We conclude that a cooling flow can be induced by Compton cooling in
the UV luminous quasar case,
since in the clustering scenario of formation of cosmic structure quasars
are expected to be formed 
in highly dense massive self-gravitating gas clouds
(Katz {\it et al.} 1994).
\hfill\break
\hfill\break

\noindent{ACKNOWLEDGEMENTS}\\
 We would like to thank Professor S. Sakasita for his continuous
encouragement and anonymous referee for the fruitful comments.
The calculations were carried out on HP715/100.
\par

\clearpage
\pagestyle{empty}
\begin{figure*}
\vspace{1.0cm}
\caption{the Compton cooling time for
$L_{\rm {UV}}=10^{46} \; \rm {erg \: s}^{-1}$ and
$L_{\rm {UV}}=10^{47} \; \rm {erg \: s}^{-1}$ 
and the radiative cooling time for initial gas with various temperature;
Initial gas distribution is given by equation (12).}
\label{fig-cool1}
\end{figure*}
\begin{figure*}
\vspace{1.0cm}
\caption{Evolution for model 7.d747. (a), (b), (c) and (d) are
         the results of distribution of density, velocity,
         temperature and accretion rate, respectively.
         In (a), the density distribution becomes more steep with time.
         In the central region, the density increases from
         $6 \times 10^{-26} \; \rm {g \: cm^{-3}}$ to
         $5 \times 10^{-24} \; \rm {g \: cm^{-3}}$.
         Since the gas concentrates, the hot gas flows inward
         from the outer region.
         As shown in (c), in the central region, the temperature decreases
         from $3 \times 10^7$ K to $1 \times 10^6$ K by Compton cooling.
         Then the flow velocity increases to
         $2.4 \times 10^7 \; \rm {cm \: s^{-1}}$
         and the accretion rate also increases to
         $0.4 \: M_{\odot} \: \rm {yr^{-1}}$
         in the central region.
         The gas temperature declines gradually, but when the radiative cooling
         time becomes shorter than the Compton cooling time,
         the gas temperature 
         declines rapidly.
         From $10^8$ yr to $10^9$ yr, the gas flow changes into the quasi-
         hydrostatic state.}
\label{fig-flow1}
\end{figure*}
\begin{figure*}
\vspace{1.0cm}
\caption{The same as Fig. 2 , but for model 7.d746
         As shown in (a) and (c), at $10^8$ yr
         the density increases and the temperature decreases,
         but the profiles
         are not more rapid than those in Fig. 2(a) and (c).
         From $10^8$ yr to $10^9$ yr, the gas state returns to the initial
         hydrostatic state.
         For this case, the gas temperature is about $2 \times 10^7$ K
         at $R_{\rm c}$.}
\label{fig-flow2}
\end{figure*}
\begin{figure*}
\vspace{1.0cm}
\caption{The same as Fig. 2, but for model 5.d747.}
\label{fig-flow3}
\end{figure*}
\begin{figure*}
\vspace{1.0cm}
\caption{The same as Fig. 2, but for model 5.d7rad.}
\label{fig-flow4}
\end{figure*}
\begin{figure*}
\vspace{1.0cm}
\caption{The same as Fig. 2, but for model 5.d746.}
\label{fig-flow5}
\end{figure*}
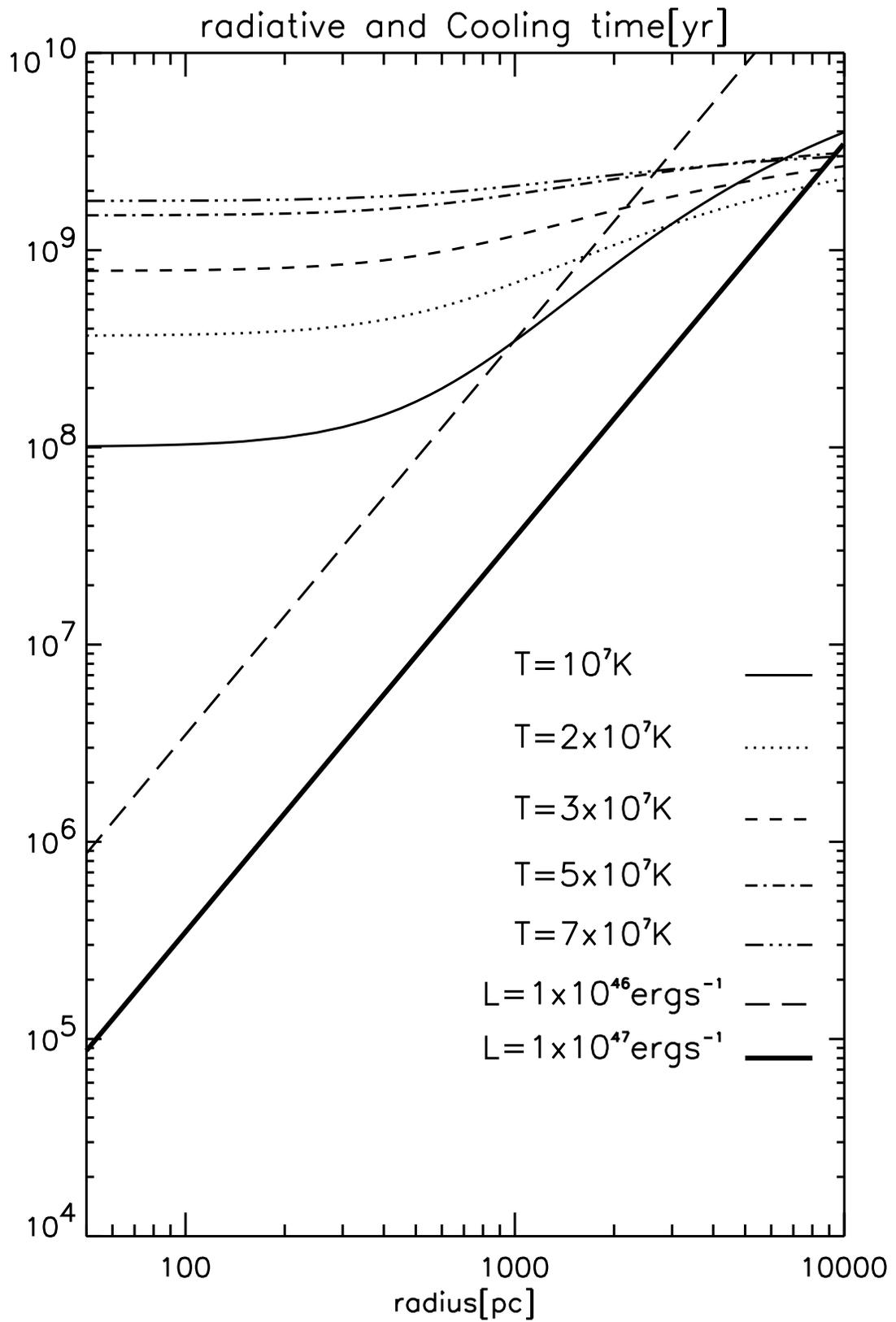
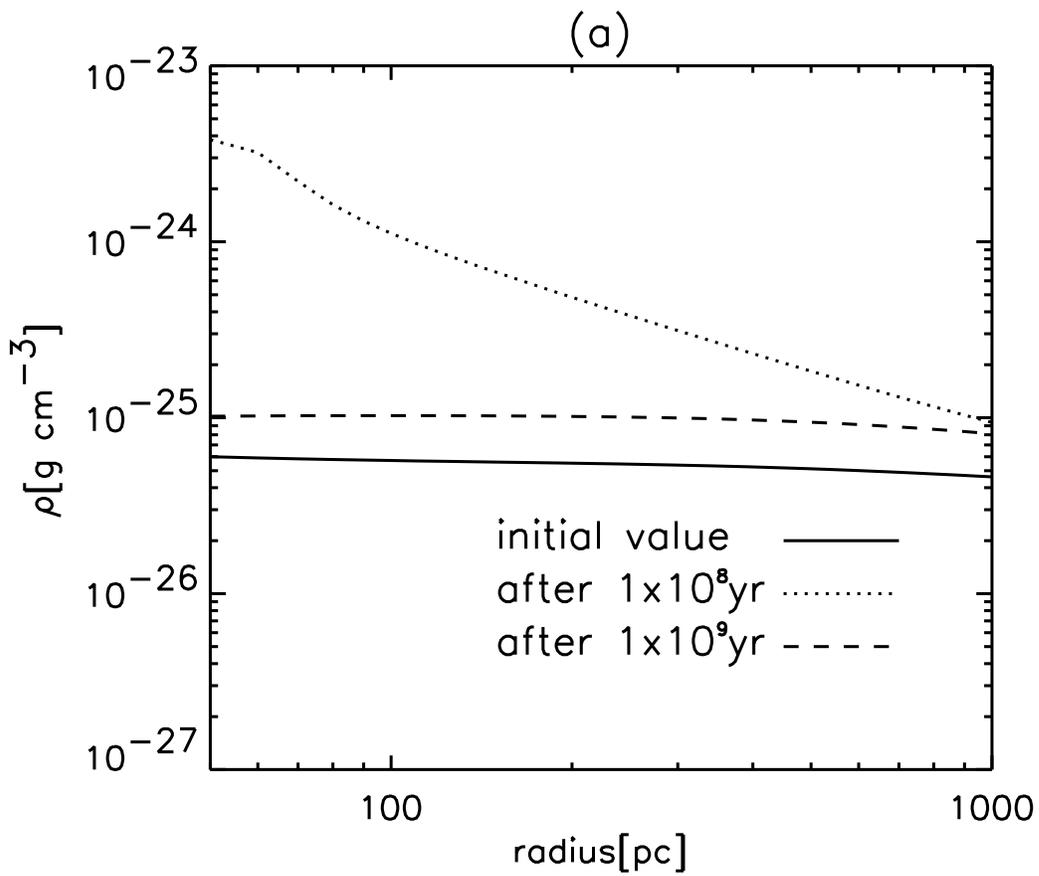
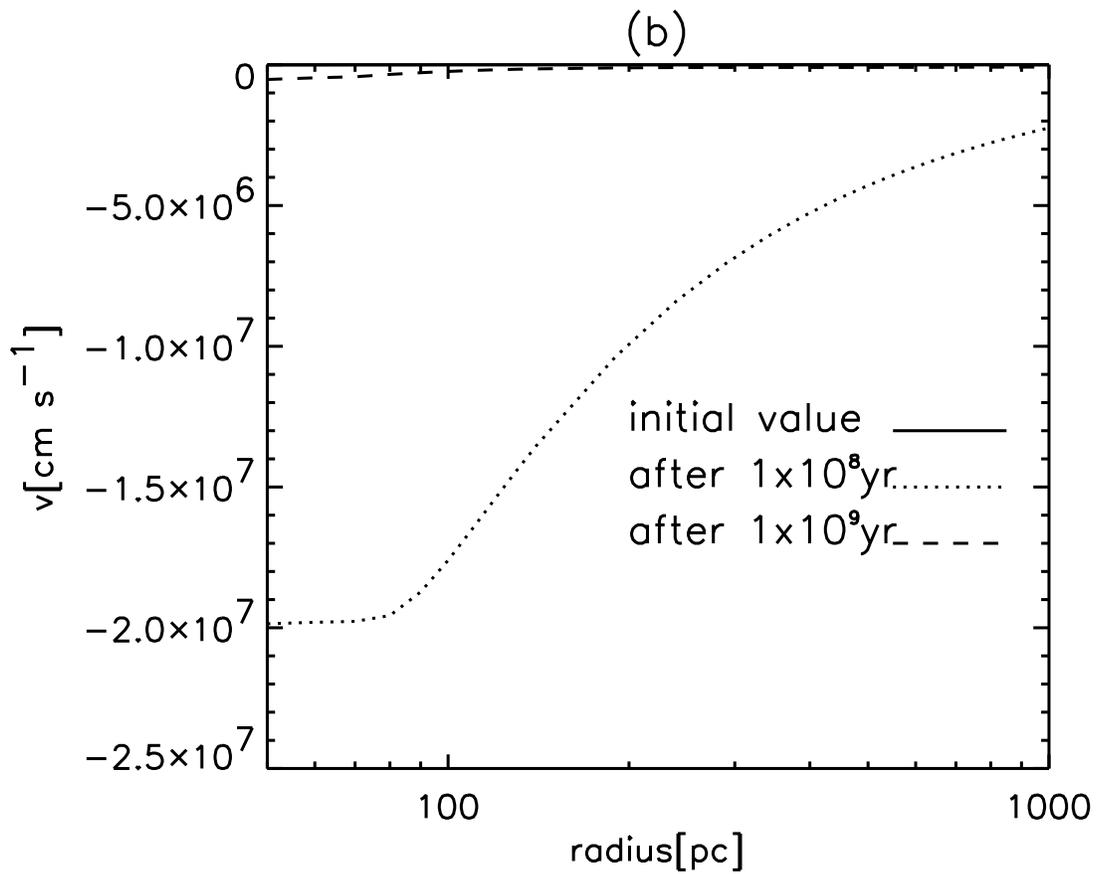
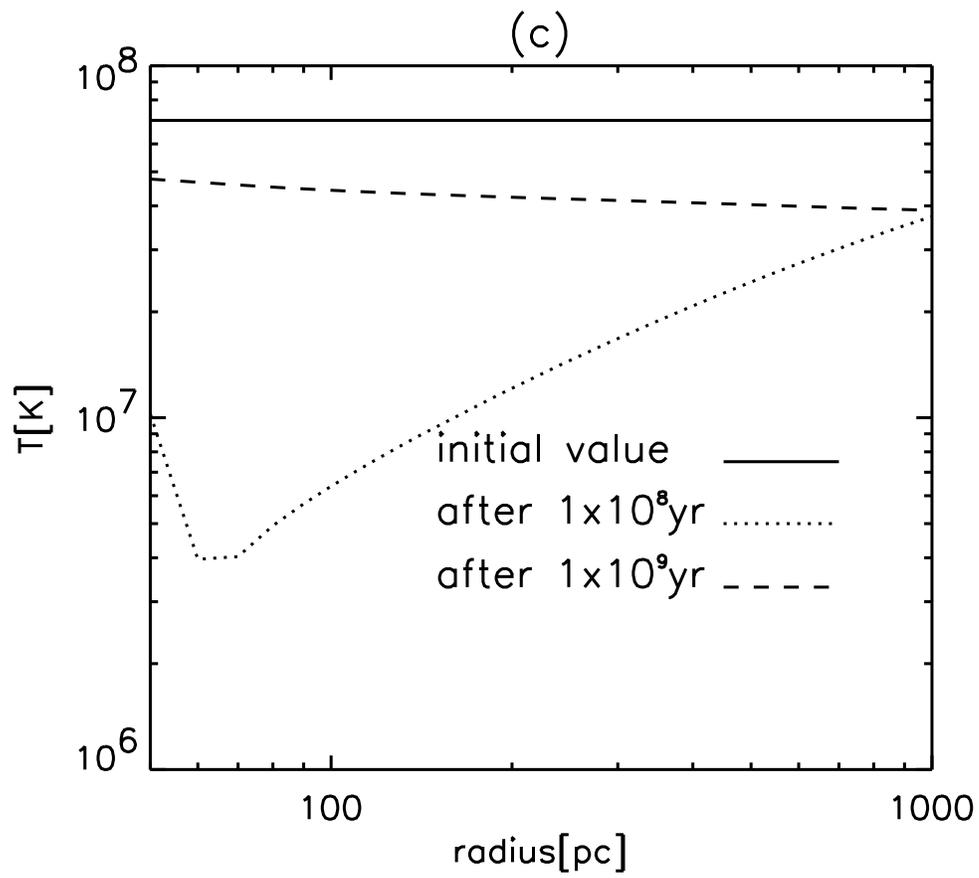
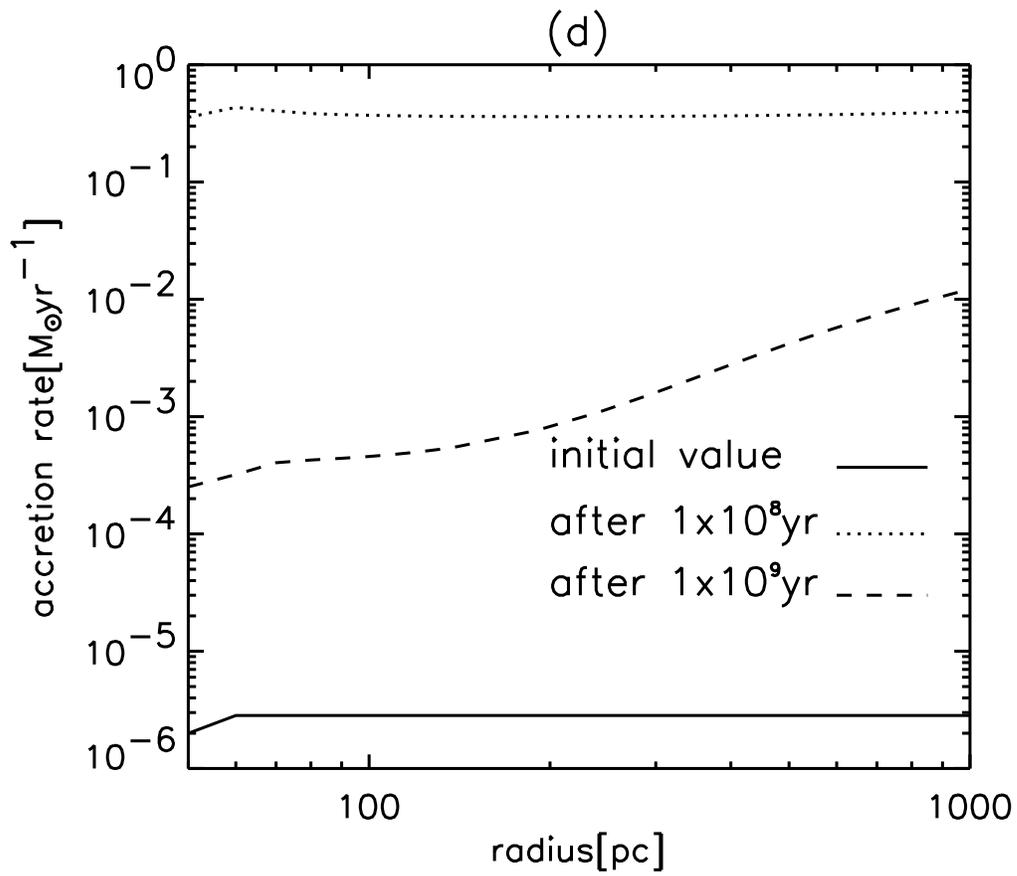
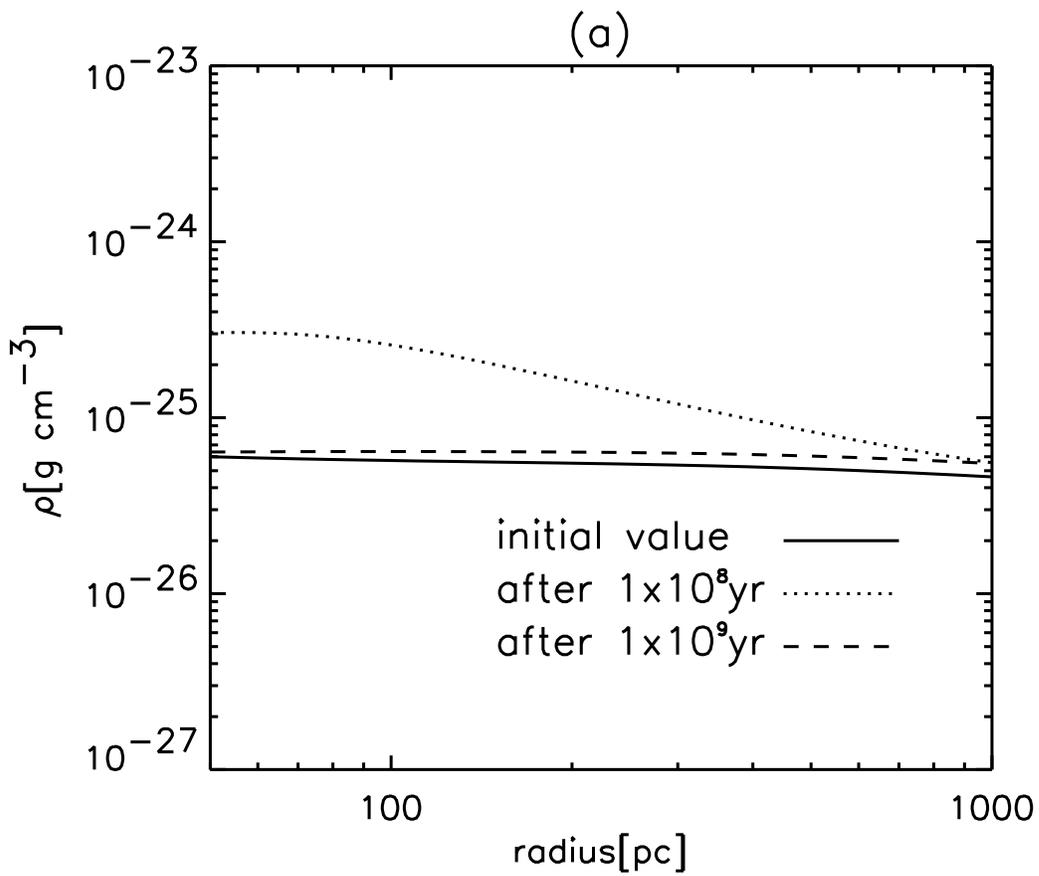
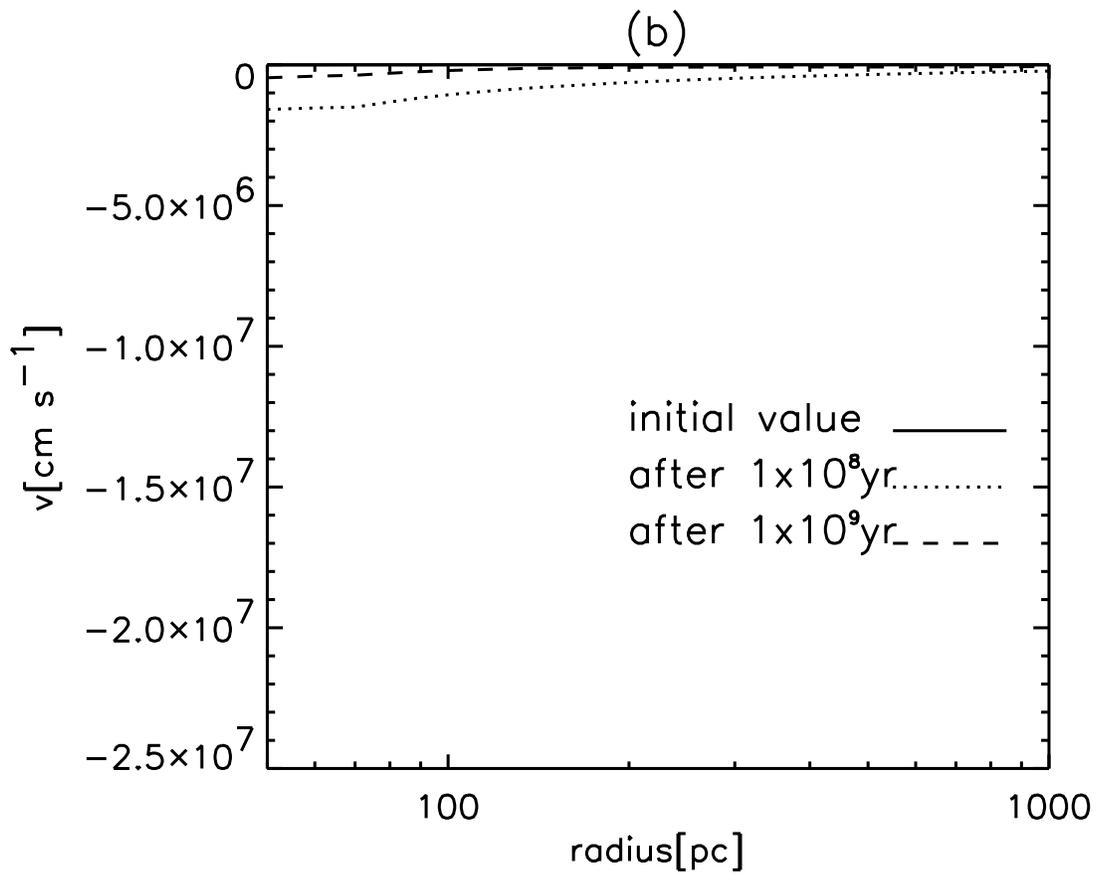
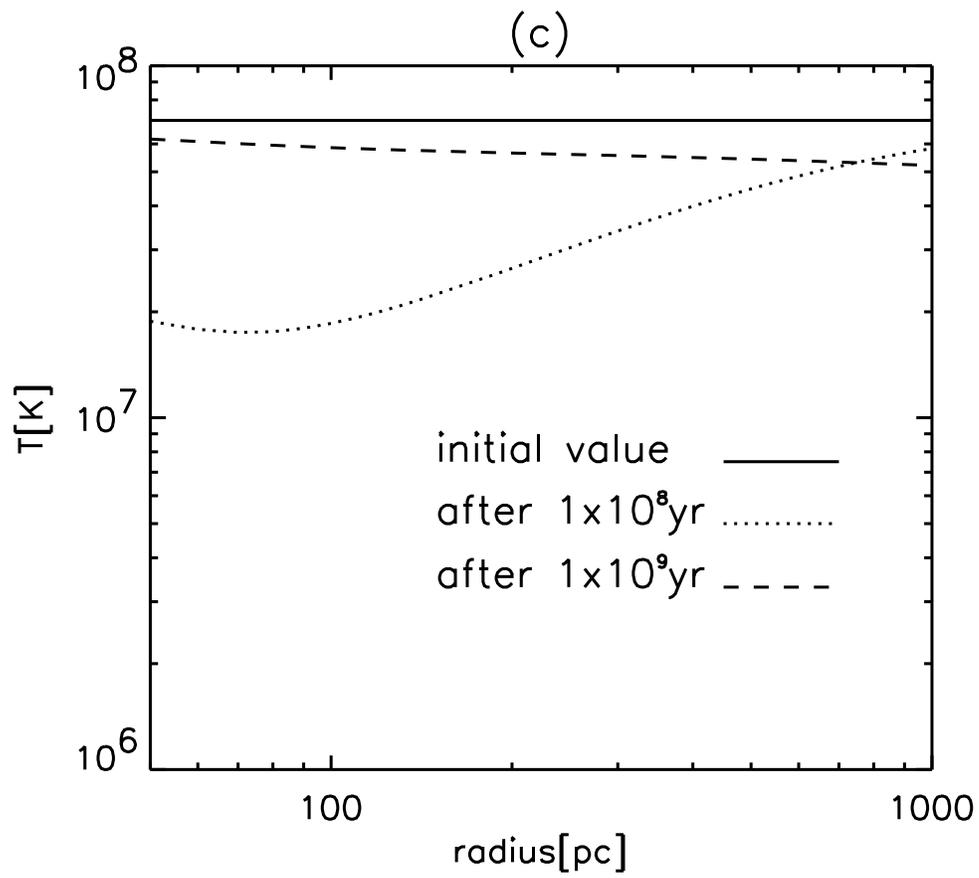
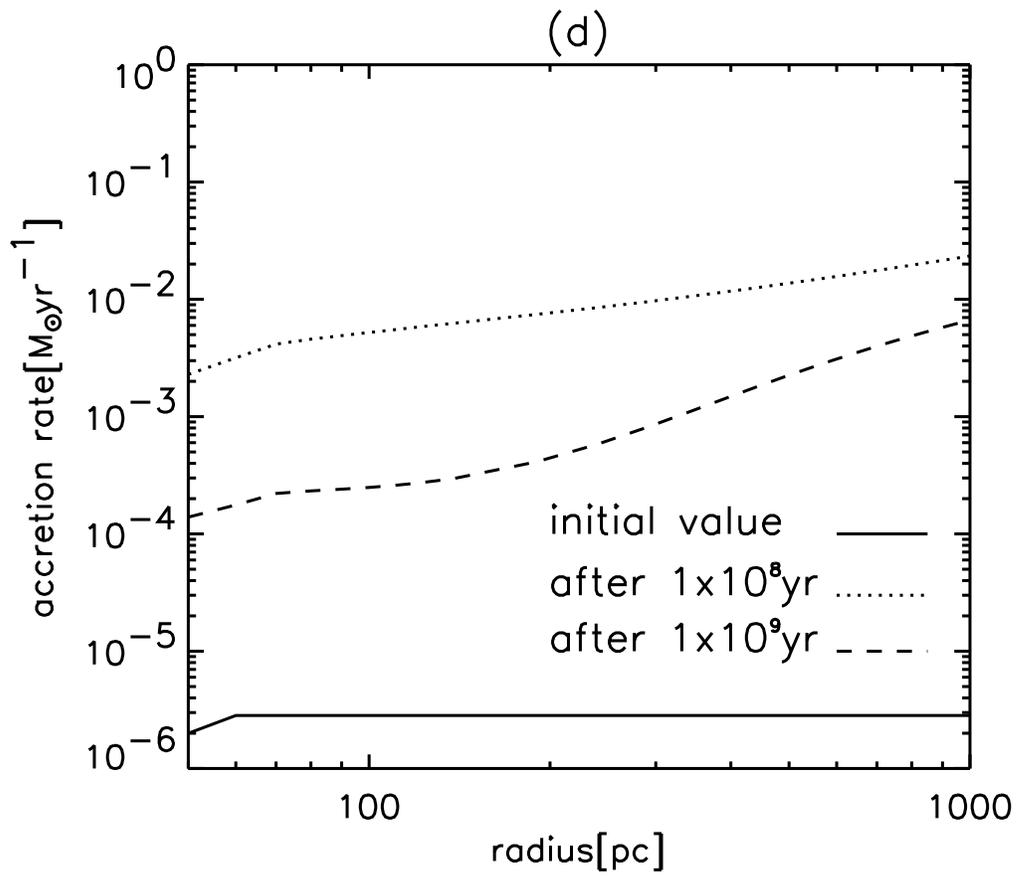
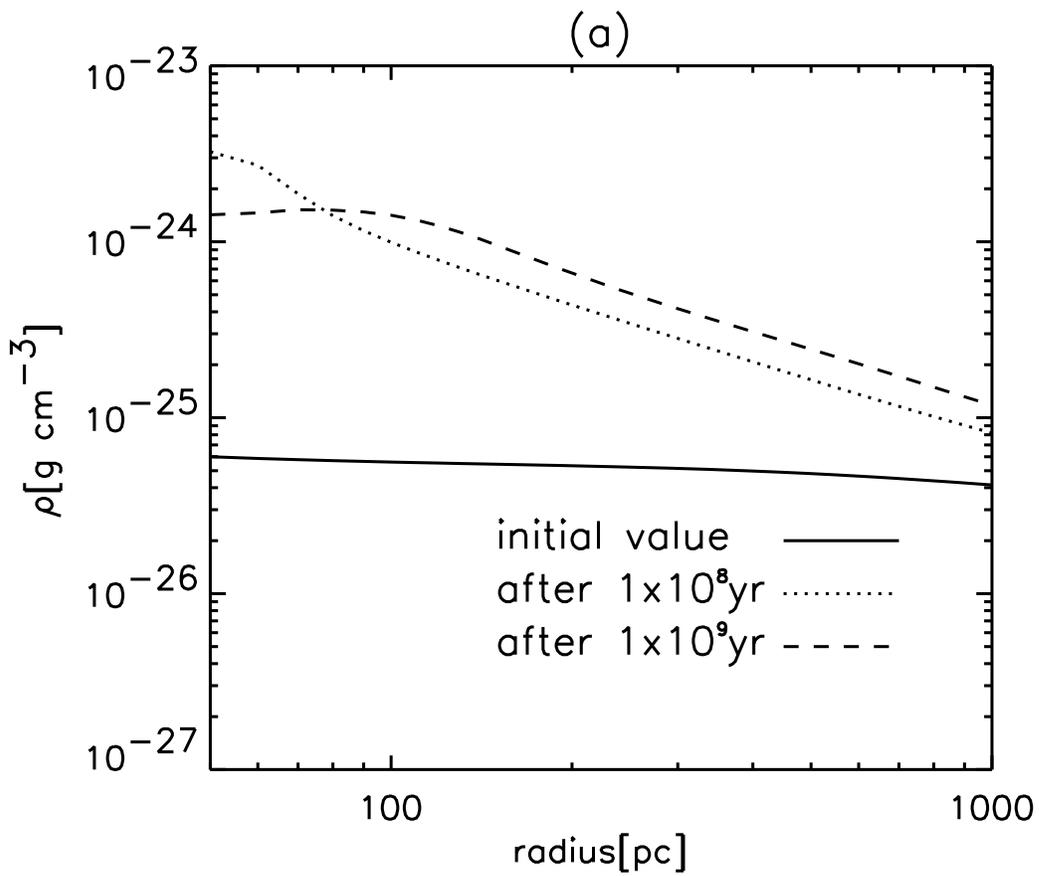
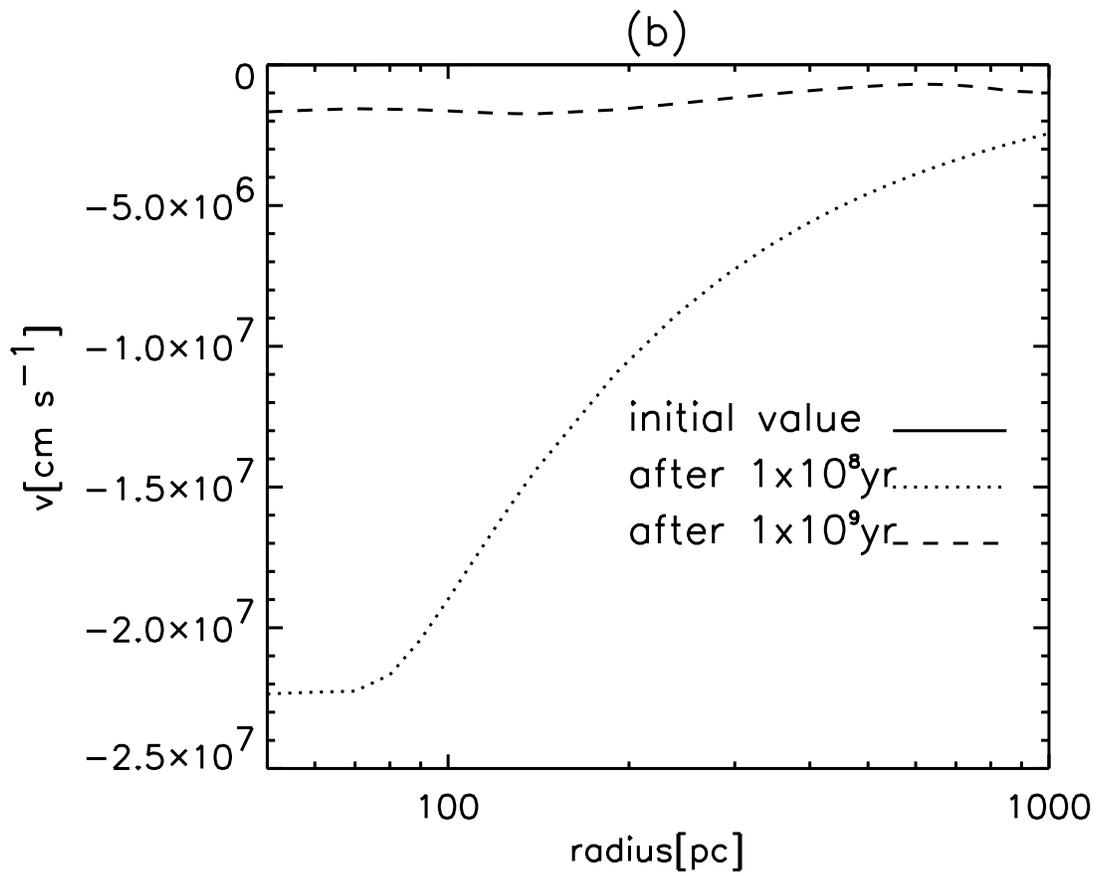
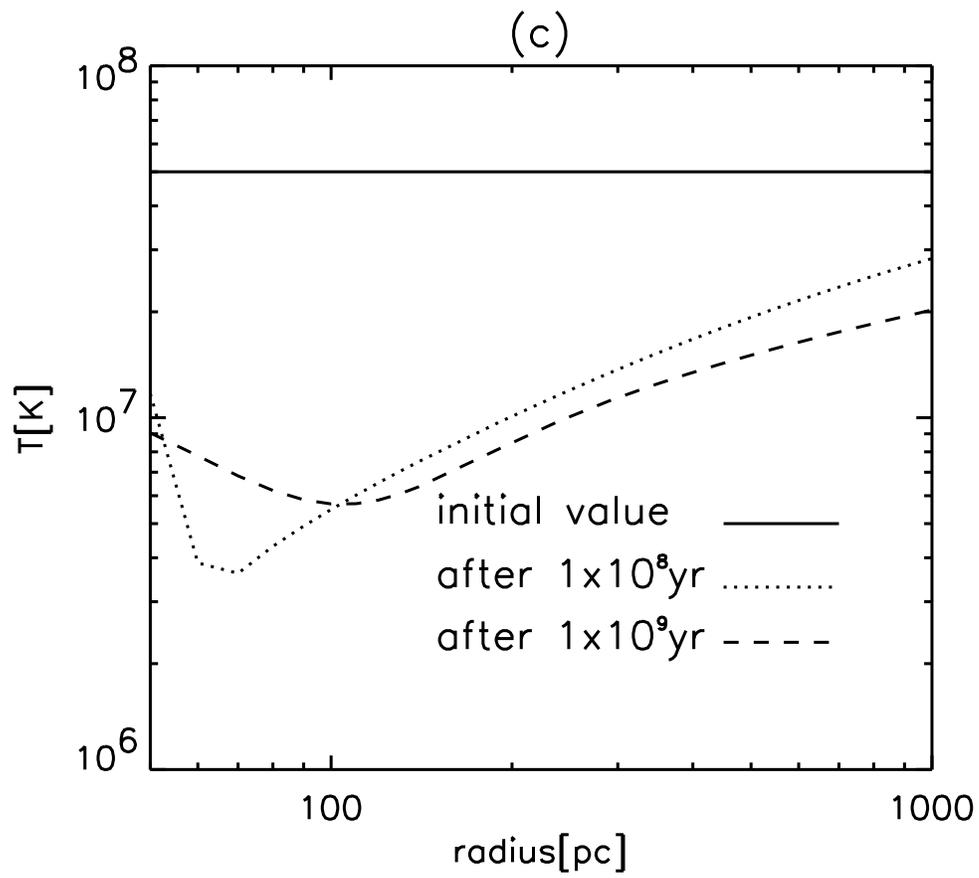
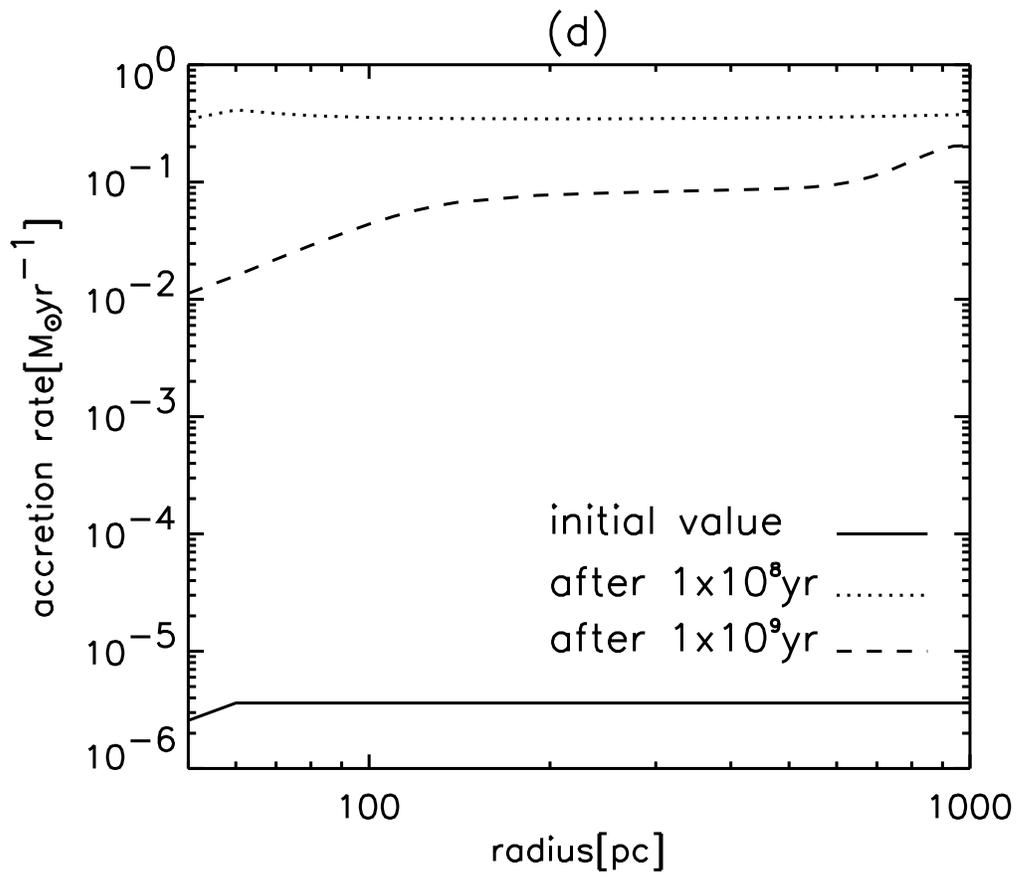
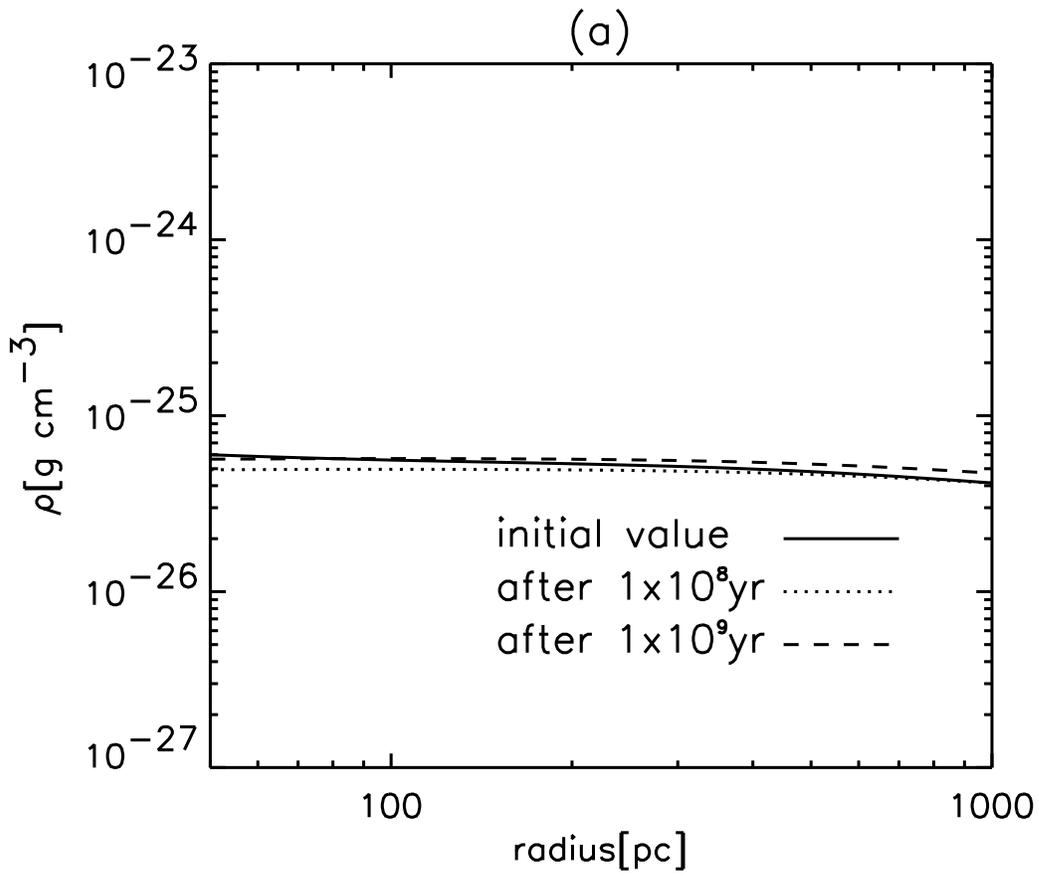
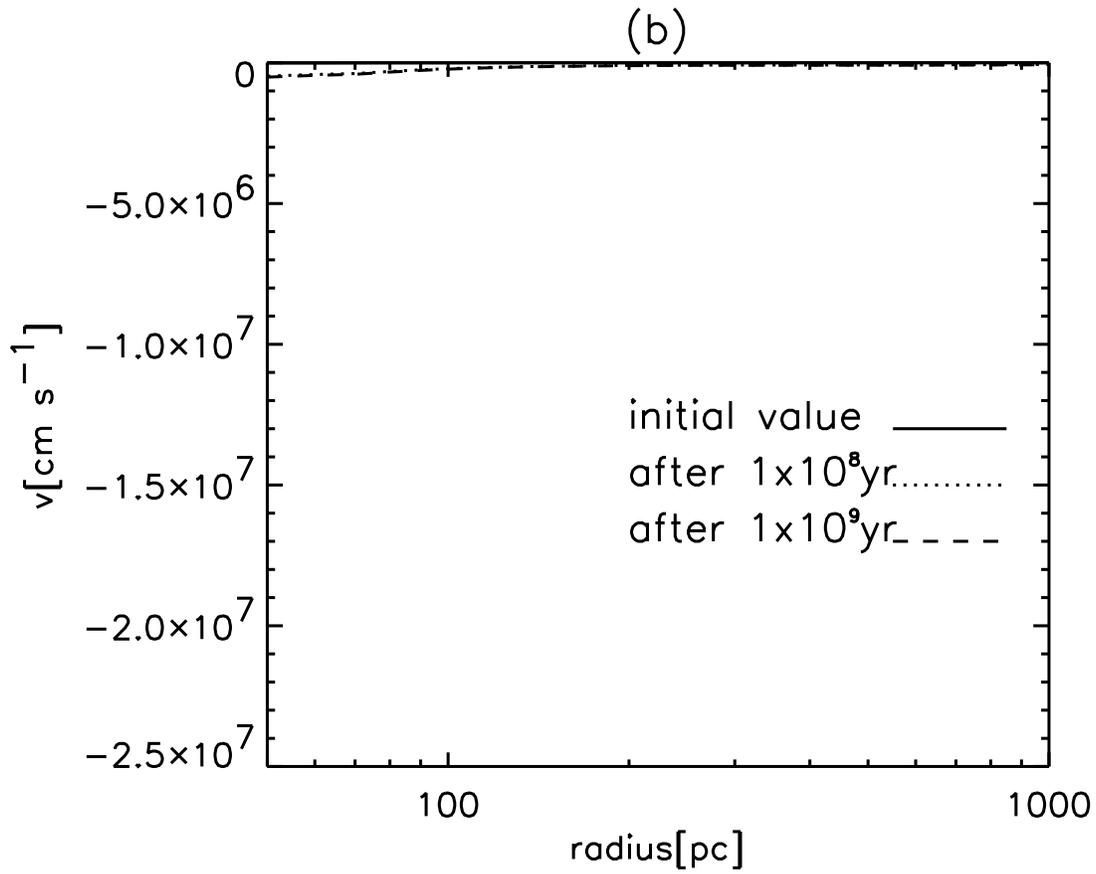
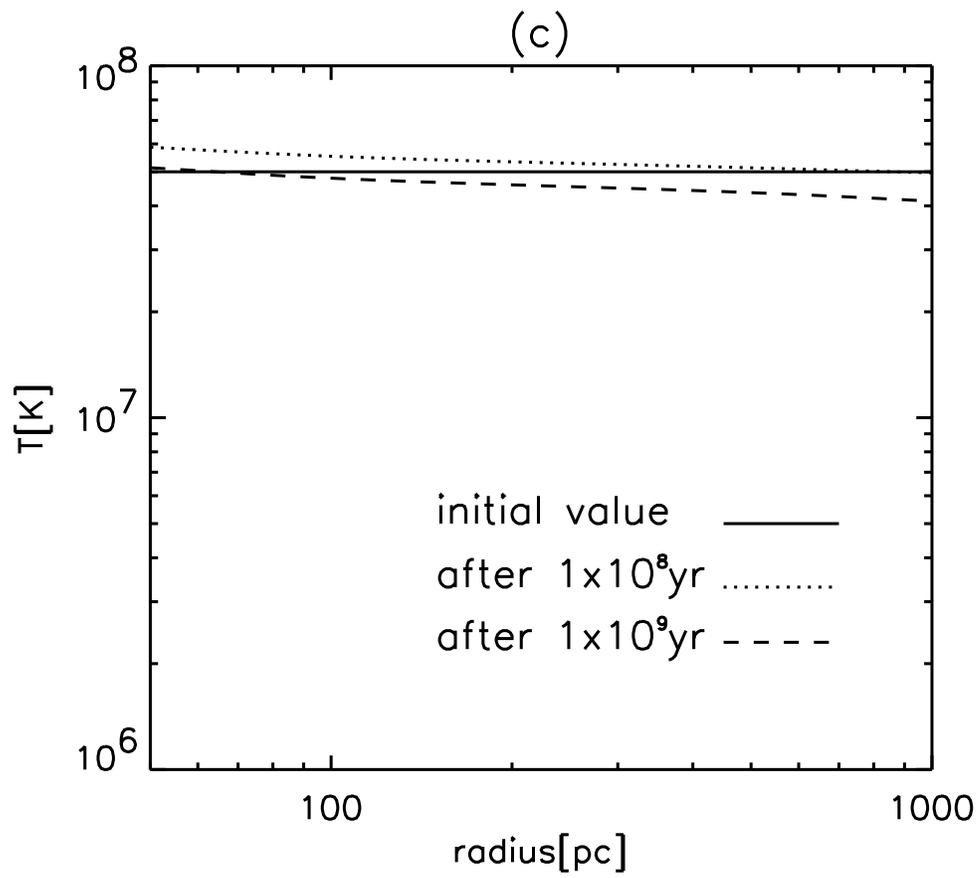
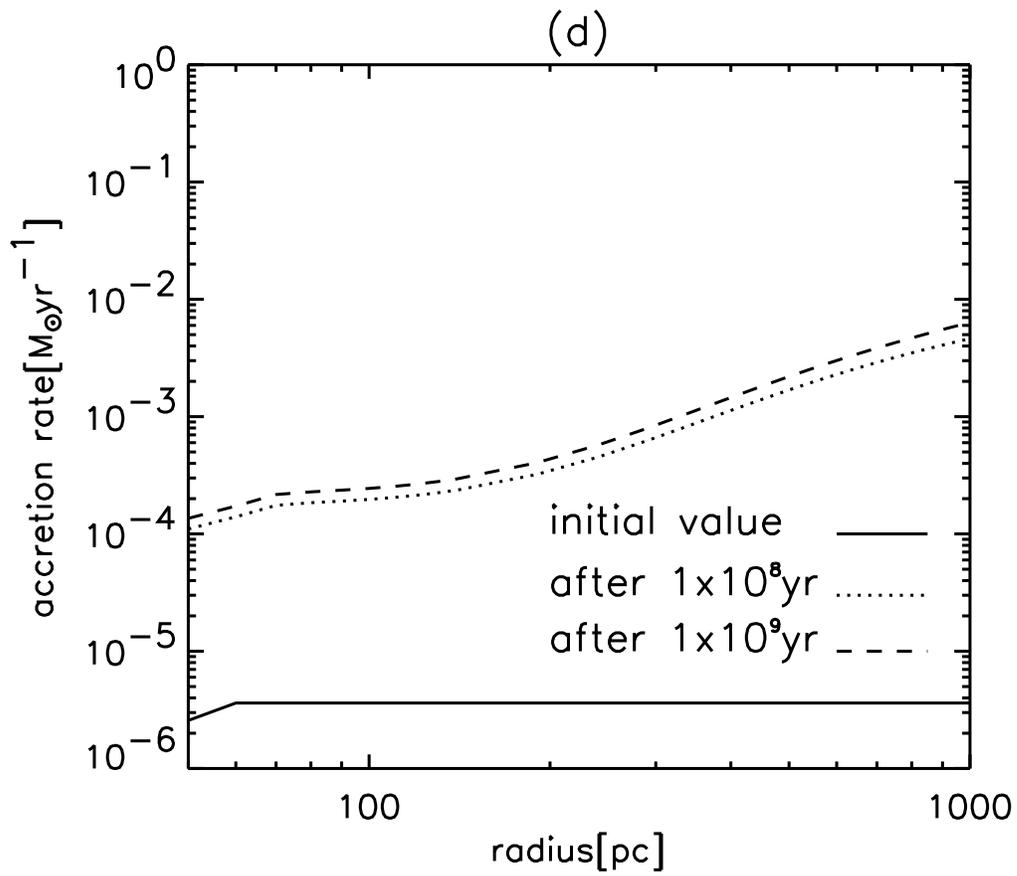
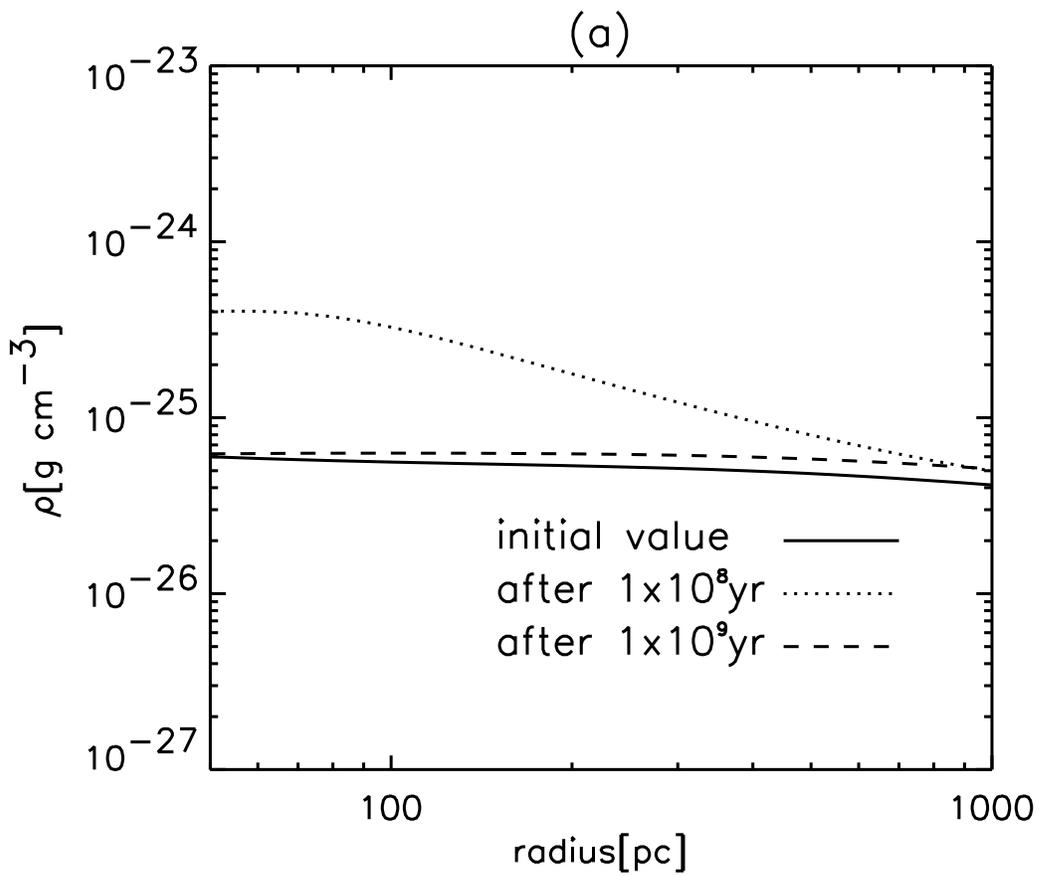
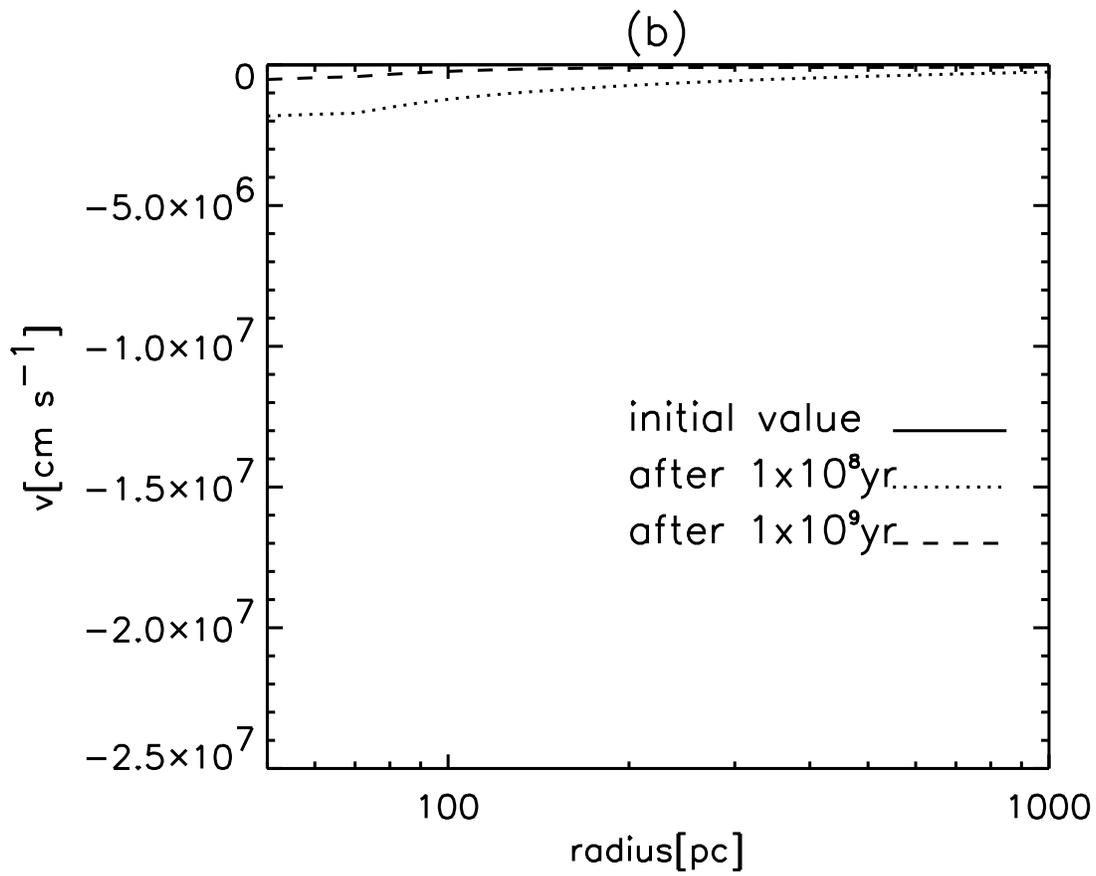
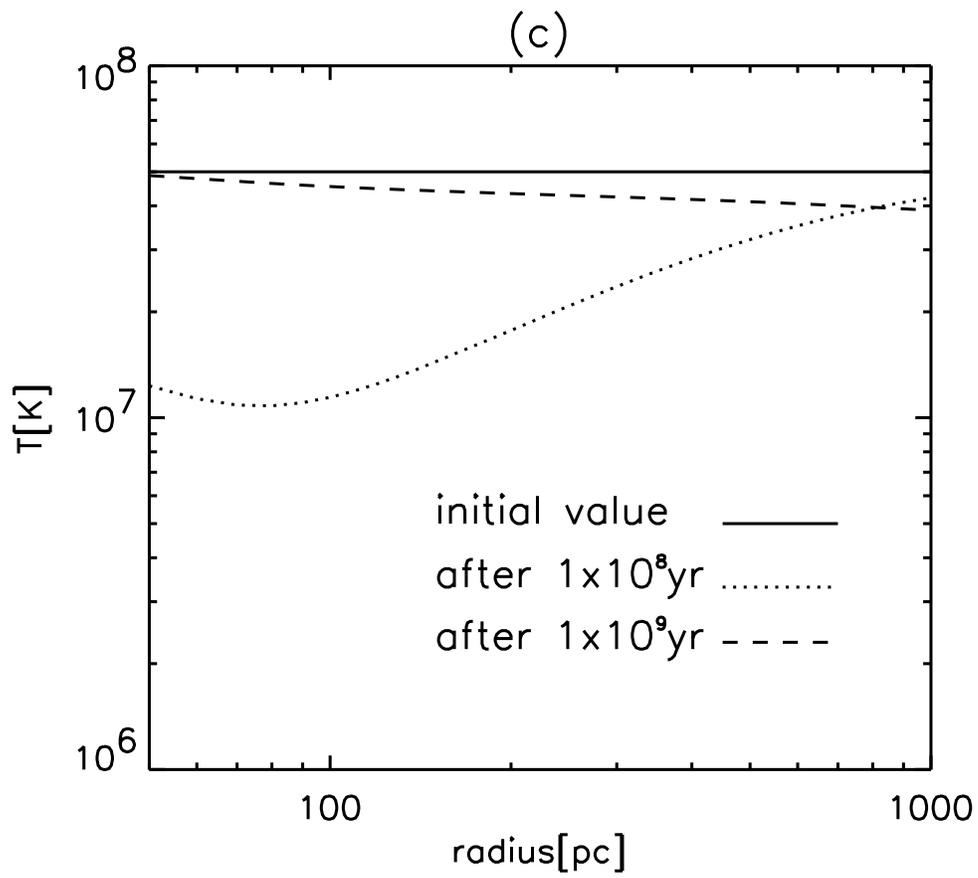
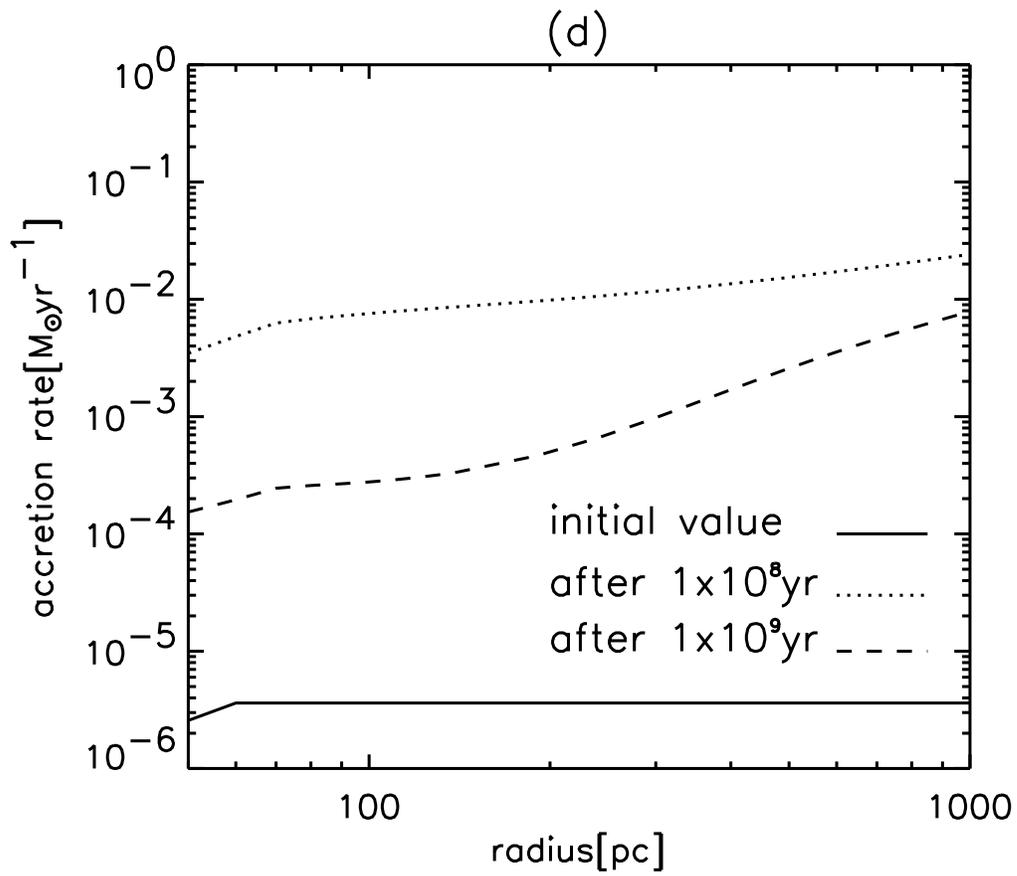
\end{document}